\documentclass[aop,apl,reprint,groupedaddress,showkeys,showpacs]{revtex4-1}

\usepackage{graphicx}

\begin{document}


\title{DC voltage-sustained self-oscillation of a nano-mechanical electron shuttle}


\author{D.~R. Koenig}
\author{E.~M. Weig}
\email{weig@lmu.de}
\affiliation{Center for NanoScience \& Fakult\"{a}t f\"{u}r Physik, Ludwig-Maximilians-Universit\"{a}t, Geschwister-Scholl-Platz 1, 80539 M\"{u}nchen, Germany}


\date{\today}

\begin{abstract}
One core challenge of nanoelectromechanical systems (NEMS) is their efficient actuation. A promising concept superseding resonant driving is self-oscillation. Here we demonstrate voltage-sustained self-oscillation of a nanomechanical charge shuttle. Stable transport at $4.2$\,K is observed for billions of shuttling cycles, giving rise to ohmic current-voltage curves with a sharp dissipation threshold. With only a few nanowatts of input energy the presented scheme is suitable for operation in the millikelvin regime where Coulomb blockade-controlled single electron shuttling is anticipated.
\end{abstract}


\maketitle


Self-oscillation, the generation of a periodic oscillation from a constant input signal in the absence of external modulated driving forces, is a well-known phenomenon in physics~\cite{bib:Pechenkin2002}. The underlying concept is based on the paradigm that even a damped resonator can oscillate continuously without periodic external driving~\cite{bib:Andronov1928}.
The required energy to overcome the dissipation and to sustain the oscillation must therefore be extracted from a constant source. This is enabled by an internal feedback mechanism regulating the energy supplied to the system per half-period.

The generic example of a self-oscillating system is the pendulum clock invented by Christian Huyghens in 1658~\cite{bib:Huyghens1658}. Other self-oscillatory phenomena in everyday life include aeroelastic galloping of iced-up overhead power lines or flutter of suspension bridges, the most famous example causing the collapse of the Tacoma Narrows Bridge in 1940~\cite{bib:Billah1991}. Both the human voice~\cite{bib:Titze1988} and the sound of a violin~\cite{bib:Popp1990} arise from mechanical self-oscillation. Similarly, self-sustained oscillations occur in many biological systems and biochemical processes~\cite{bib:Pol1928,bib:Novak2008}, controlling e.g. the beating of the heart or circadian cycles in body temperature.

\begin{figure}
\includegraphics[width=8.5cm]{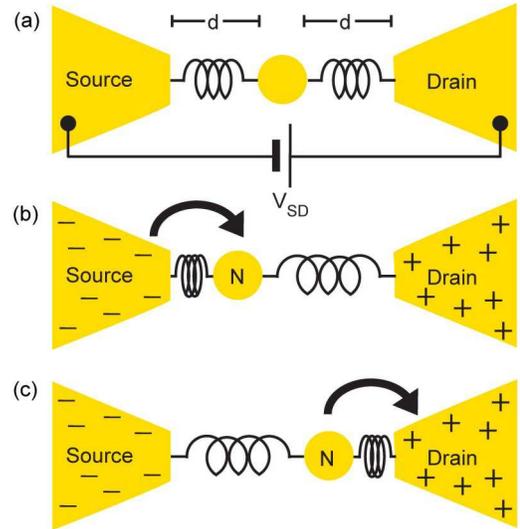}\\
\label{figure1}
\caption{Concept of mechanical charge
transport. (a) Idle shuttle at rest position.
(b) and (c) Operating shuttle being charged/uncharged with $N$ electrons at the source/drain contact, respectively.}
\end{figure}

\begin{figure*}
\includegraphics[width=17cm]{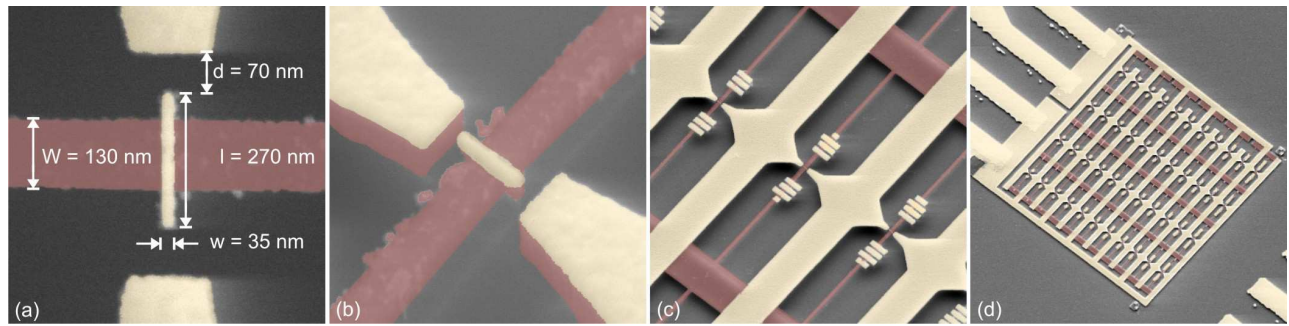}\\
\label{figure2}
\caption{The nanomechanical electron
shuttle. (a) SEM image of a shuttle indicating the dimensions of gold island (yellow) and high stress silicon nitride resonator (red) suspended above the silicon substrate (grey). (b) - (d) Tilted view of gold island between source and drain electrode, entire resonator including custom set of gold weights and clamping points, as well as zoomed out perspective of electrical contacts shunting an array of forty four shuttling devices, respectively.}
\end{figure*}

In the physics realm, the Franklin bell~\cite{bib:Franklin1753,bib:Tuominen1999} or electrical devices such as the van der Pol oscillator~\cite{bib:Pol1920}
and the Gunn diode~\cite{bib:Gunn1963} are landmark self-oscillating systems.
Mathematically, the dynamics of this kind of nonlinear system is analyzed in 2D phase space. While the solutions of conservative systems consist of fixed points or closed orbits reminiscent of stable equilibrium positions or cyclic trajectories, respectively, the situation is more complex for nonconservative systems. As soon as dissipation enters, a third type of solution, limit cycles, have to be considered. Limit cycles describe isolated closed trajectories which attract adjacent ones, forming so-called basins of attraction in phase space~\cite{bib:Strogatz1994}. The existence of a stable limit cycle implies a periodic solution of the system giving rise to self-sustained oscillation in the absence of external periodic forcing.
Its amplitude and frequency are largely independent of the initial conditions, such that the periodic trajectory is stable towards small external disturbances.

Consequently, self-oscillation allows to convert a direct current (DC) input into a stable oscillation, which makes it a powerful transduction mechanism for mechanical systems. In particular, the actuation by means of self-oscillation is a viable option for micro- and nanomechanical systems~\cite{bib:Ekinci2005} where the quest for efficient, non-dissipative driving schemes is ongoing.
We would like to note that similarly, external feedback can be employed to obtain self-sustaining oscillation of a nanomechanical system, in that case consisting of a resonator as well as an external oscillator, as e.g. in~\citeauthor{bib:Feng2008}~\cite{bib:Feng2008}.
However this sets the phase of the resonator and is thus conceptionally different from auto-oscillating systems which are in the focus of the present work.

In recent years, various schemes of self-oscillation have been employed to efficiently actuate nanomechanical resonators. Optomechanical systems can be driven by bolometric~\cite{bib:Aubin2004,bib:Metzger2008a} or radiation pressure feedback~\cite{bib:Kippenberg2005,bib:Anetsberger2009} which equally applies in the microwave domain~\cite{bib:Faust2012a}.
In nanoelectromechanical systems, internal feedback has been realized by field emission of vibrating nanowires subject to a DC voltage~\cite{bib:Ayari2007,bib:Kim2010b,bib:Weldon2010},
the periodic charging of a nanowire in the constant electron beam of a scanning electron microscope~\cite{bib:Vincent2007}, or transport through a carbon nanotube quantum dot mediated by the backaction of tunneling single electrons~\cite{bib:Steele2009a}. Recently, the thermodynamic feedback of a piezoresistive resonator~\cite{bib:Steeneken2011} or optical band-gap excitation in a GaAs heterostructure cantilever~\cite{bib:Okamoto2011a} have been employed to achieve self-oscillation.

A particularly striking example is voltage-sustained self-oscillation in a nanomechanical charge shuttle which has been proposed by Gorelik {\it et al.} in 1998~\cite{bib:Gorelik1998}. In this system a nanoscale metallic island hosted by a nanomechanical resonator can oscillate between a nearby source and drain electrode biased with a DC voltage
$V_{\mathrm{DC}}$ as depicted in Fig.~1(a). When mechanically excited, the island can pick up $N$ electrons at the source
electrode and mechanically transfer them to the drain (Fig.~1(b),(c)). This generates a modulated DC current current which amounts to $\langle I_{\mathrm{SD}} \rangle = 2e\langle N \rangle f$ where $e$ is the elementary charge, $N$ is the number of excess charge carriers transported per half-period, $\langle N \rangle$ is its thermal average and $f$ is the mechanical eigenfrequency.

At the same time, the electric field between the plates exerts a force on the charged island which accelerates the shuttle.
In a simple parallel plate capacitor model assuming negligible screening the electric field can be written as $E = V_{\mathrm{DC}}/(2d)$ with symmetric island-electrode separation $d$ (see Fig.~1(a)).
Above a certain threshold voltage $V_{\mathrm{th}}$, the electrostatic energy $U_{\mathrm{el}} = 2 eN \cdot E \cdot 2d = 2eN V_{\mathrm{DC}}$ overcompensates the mechanical dissipation $U_{\mathrm{diss}}$ of the system such that self-oscillation based on the repetitive charge reversal at the electrodes in the static electric field has been predicted~\cite{bib:Gorelik1998,bib:Rueting2005,bib:Shekhter2006a} much like in the shuttle's macroscopic counterparts~\cite{bib:Franklin1753,bib:Tuominen1999}.

Here we present a nanomechanical charge shuttle operated solely by an applied DC bias voltage. Previous experiments on charge shuttling have mostly relied on external actuation to enable charge transfer between source and drain~\cite{bib:Erbe2001,bib:Scheible2004,bib:Koenig2008}.
Specifically, electrically applied RF signals have been widely employed and led to significant experimental advancements in the field~\cite{bib:Erbe2001,bib:Scheible2004}.

However, the application of large RF voltages fundamentally limits the shuttle performance due to undesired interactions with the charged island.
This constraint has been resolved by the implementation of an acoustically driven shuttle, inertially actuated by means of ultrasonic waves~\cite{bib:Koenig2008}. On the other hand, the dissipation of the required piezo transducer gives rise to substantial heating of the system, inhibiting operation at or below $4$\,K. The above limitations have so far been a major obstacle for observing single electron shuttling in the Coulomb blockade regime~\cite{bib:Gorelik1998}.
Shuttle realizations reporting voltage-sustained self-oscillation~\cite{bib:Moskalenko2009,bib:Kim2010b}
have operated in a regime of extremely small mechanical amplitudes
and have not been able to yield ohmic response as expected from a moveable single electron box in the high temperature regime.

The nano-mechanical electron shuttle under investigation depicted in Fig.~2 consists of a gold island with typical dimensions of $w \times l \times h = 35$\,nm $\times 270$\,nm $\times 40$\,nm. The island is placed in the center of a doubly clamped freely suspended silicon nitride string, which is $L = 14\,\mu$m long, $W =130$\,nm wide and $H = 100$\,nm high. We employ high stress LPCVD-grown silicon nitride incorporating an intrinsic tensile stress of $1.38$\,GPa~\cite{bib:Unterreithmeier2009}, a material which
exhibits strong restoring forces, preventing stiction of the island to the side electrodes. The latter are placed symmetrically on either side, leaving a $d =70$\,nm gap to the island. In order to obtain a large device density allowing for statistically representative results a highly parallelized approach is chosen: Forty four shuttling devices are electrically shunted between two interdigitated comb electrodes, and selectively addressed via frequency multiplexing realized by sets of custom gold weights attached to each individual resonator. We would like to note that the number of addressable devices is highly scalable and that prototypes with several hundreds of shuttles have been fabricated.

Our experiments are performed in helium exchange gas with $p = 0.5$\,mbar in a helium dewar at $T = 4.2$\,K. Several shuttle chips with slightly varying device dimensions have been investigated. While voltage-sustained self-oscillation has been observed in several devices, the results shown in this work (except Fig.~4(c)) are from one
representative array. The time-averaged DC current $\langle I_{\mathrm{SD}} \rangle = 2e\langle N \rangle f$ is measured with a low noise current preamplifier. Inertial actuation mediated by a piezo actuator is employed to characterize the shuttle eigenfrequencies and response in the driven shuttling regime as described in detail in~\citeauthor{bib:Koenig2008}~\cite{bib:Koenig2008}.

In order to observe voltage-sustained self-oscillation, the resonant drive is switched off after an initial trigger required to charge the island with $\pm Ne$ at the source or drain contact.
Self-sustained shuttling is subsequently maintained by the electric field $E = V_{\mathrm{DC}}/(2d)$ created between the two voltage-biased electrodes.
It will give rise to an electrostatic force $F_{\mathrm{el}} = Ne \cdot E$ accelerating the island charged with $N$ electrons towards the oppositely charged electrode. Upon contact, the island charge is reversed to $-Ne$, leading to a sign change of the force $F_{\mathrm{el}} = - Ne \cdot E$ and a subsequent acceleration of the island back to the initial electrode. Thus the internal feedback mechanism required for self-oscillation is provided by periodic charge reversal of the island with $2f$.

\begin{figure}
\includegraphics[width=8.5cm]{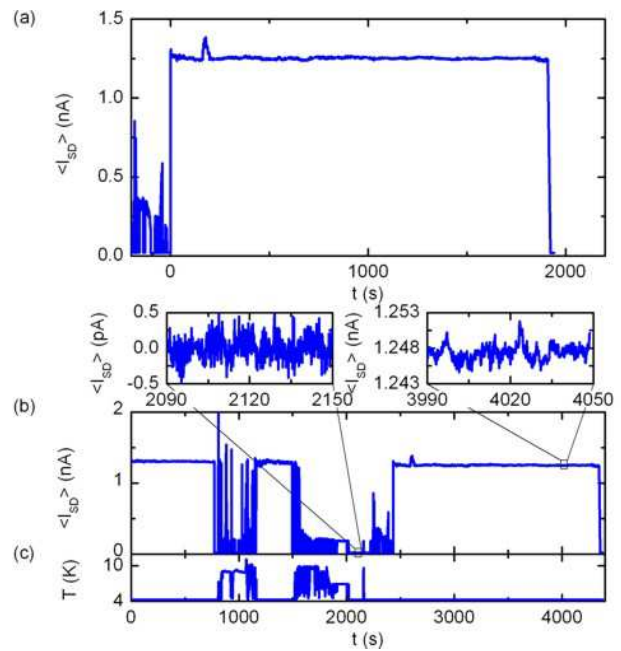}\\
\label{figure3}
\caption{Self-oscillation and current stability.
 (a) Shuttling current as a function of time displaying voltage-sustained self-oscillation at $V_{{\mathrm{SD}}} = 12$\,V.
  (b)Shuttling current as a function of time showing multiple events of voltage-sustained self-oscillation separated by regimes of acoustically driven shuttling with varying actuation frequency and power.
 Insets compare the noise floor of the experimental setup in the idle case without nanomechanical motion and the stability of the shuttling current in the self-oscillation regime.
 (c) Sample temperature as a function of time indicating regimes of self-oscillation and strongly driven shuttling.}
\end{figure}

Figure~3(a) shows for $V_{{\mathrm{SD}}} = 12$\,V how current transport sets in at $t=0$\,s after a short acoustic trigger of $30$\,dBm applied at the resonance frequency $8.99$\,MHz and remains unchanged after the resonant actuation has been turned off. Note that the current spike near $t=200$\,s is a calibration peak that has been applied in order to characterize the impedance-dependent offset of the voltage amplifier. Self-sustained oscillation goes on for almost $2,000$\,s, which corresponds to roughly $10^{10}$ cycles, until the shuttling current eventually collapses to zero, presumably due to impact-induced coupling to out-of-plane or torsional motion or wear-induced alteration of island and/or electrodes. This collapse of the shuttling current is not to be mistaken with a breakdown of the shuttling device: We would like to emphasize that self-oscillation can be re-established by a new trigger as shown in Fig.~3(b), albeit its initial parameters or the required bias $V_{{\mathrm{SD}}}$ might vary slightly. The strong time-dependence of the shuttling current in the time interval preceding $t=0$\,s in Fig.~3(a) or between $t=800$  and $2400$\,s in in Fig.~3(b) is a consequence of a variation of actuation frequency and power as well as bias to identify regimes of stable shuttling
capable of self-oscillation.
The transitions between piezo-driven
shuttling and voltage-sustained self-oscillation are also apparent from Fig.~3(c), where the sample temperature is plotted over time. While the temperature is increased by up to several kelvin during driven shuttling due to piezo heating, it quickly saturates at the bath temperature of $4.2$\,K during self-oscillation.

The left and right inset of Fig.~3(b) show close-ups of $60$\,s intervals of the measured current: While the noise floor measured with the idle device (left) displays RMS fluctuations of $0.2$\,pA due to amplifier noise, the RMS fluctuations of the current across the self-oscillating shuttle amount to $1.1$\,pA.
 The typical current stability of $\pm 0.1\%$ is enabled by a careful redesign of the resonator. Unlike in previous devices~\cite{bib:Koenig2008} a horizontal resonator design with $W>H$ has been chosen in order to suppress mode coupling between the in-plane shuttling mode and unwanted torsional modes of the device excited by the repetitive impact with the source/drain electrodes. This considerably stabilizes the shuttling current in comparison to previous shuttle designs with $W<H$, presumably due to a more reproducible island-electrode approach and thus charge transfer during every half-cycle. In order to further reduce mode coupling, future devices will incorporate $W \gg H$ as well as redesigned gold weights.

The average number of excess charge carriers $\langle N \rangle$ can be tuned by varying the bias voltage $V_{\mathrm{DC}}$. This is shown in Fig.~4(a), where the shuttling current is plotted for two voltage sweeps which have been taken after a $29$\,dBm/$30$\,dBm acoustic trigger with $f = 7.86$\,MHz at $V_{\mathrm{SD}}=\pm 7$\,V:
The blue trace corresponds to a reduction of $V_{\mathrm{SD}}$ from $+7$\,V to $-7$\,V, while the red trace has been taken while increasing $V_{\mathrm{SD}}$ from $-7$\,V to $+7$\,V.
The quasi-ohmic current-voltage characteristic reflects the $\langle I_{\mathrm{SD}}(V_{\mathrm{DC}})\rangle = 2 e \langle N(V_{\mathrm{DC}}) \rangle f$ behavior of a nanomechanical shuttle in the high temperature regime~\cite{bib:Koenig2008}. In addition, Fig.~4(a) shows a sharp transition to $\langle I_{\mathrm{SD}} \rangle =0$\,A at $V_{\mathrm{DC}} = 4.76$\, and $-4.48$\,V, respectively.
The threshold current of $0.5$\,nA corresponds to a minimum number of roughly $200$ electrons required to sustain self-oscillation.
The abrupt collapse of the nanomechanically transduced current is expected for the case of a damped oscillator. It occurs when the electrostatic energy $U_{\mathrm{el}} = 4 \cdot d \cdot F_{\mathrm{el}}$ provided by the DC voltage no longer exceeds the total energy $U_{\mathrm{diss}}$ dissipated per oscillation period. Thus, the threshold voltage $V_{\mathrm{th}} = U_{\mathrm{diss}}/(2eN)$ can be employed to estimate the power dissipation of the shuttle, which, in a highly nonlinear system such as the impacting shuttle is not accessible through the quality factor of the resonator. Equating
\begin{equation}
P_{\mathrm{diss}}(V_{\mathrm{th}})
\stackrel{!}{=} P_{\mathrm{el}}(V_{\mathrm{th}}) = V_{\mathrm{th}} \cdot \langle I_{\mathrm{SD}}(V_{\mathrm{th}})\rangle
\label{disspower}
\end{equation}
yields $2.4$\,nW
and $2.2$\,nW
for the blue and red curve, respectively.

A major advantage of purely DC-biased self-sustained shuttle operation is the significant decrease of the external heat load on the system. The data shown in Fig.~4 has been taken at $T = 4.2$\,K, unlike the data discussed in~\citeauthor{bib:Koenig2008}~\cite{bib:Koenig2008} where piezo heating
resulted in sample temperatures of $T>10$\,K (see also Fig.~3(c)). Furthermore, the observed power dissipation in the nanowatt range is far below the cooling power of conventional cryogenic systems. Therefore, voltage-sustained self-oscillation opens the pathway to Coulomb blockade limited electron shuttling in the millikelvin regime.

\begin{figure}
\includegraphics[width=8.5cm]{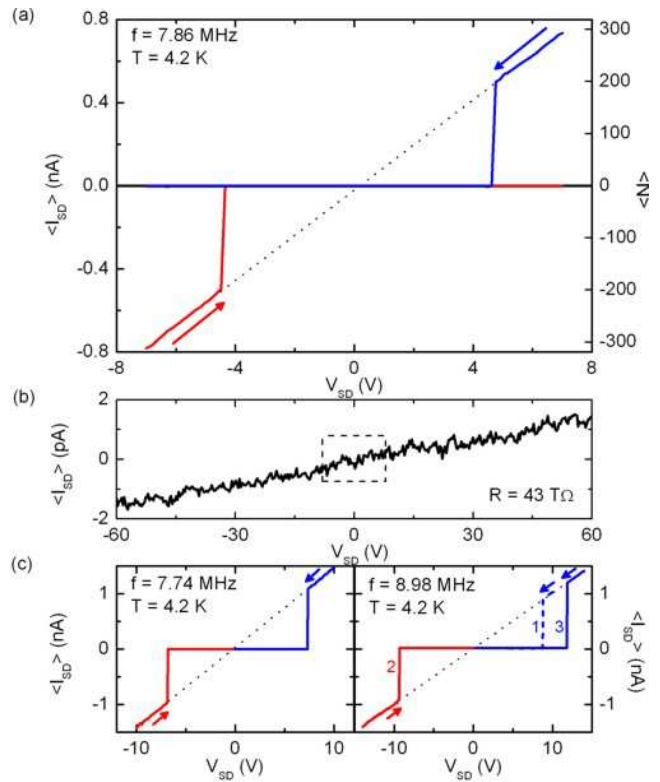}\\
\label{figure4}
\caption{DC voltage-sustained electron shuttling and background current.
(a) Current-voltage curves of voltage-sustained self-sustained oscillation. Both blue and red trace corresponding to downward and upward voltage sweep, respectively, feature a sharp dissipation threshold.
(b) Background current determined by measuring $I_{\mathrm{SD}}$ in the absence of mechanical shuttling as a function of $V_{\mathrm{SD}}$. The dashed box indicates the voltage range depicted in {\bf a}, where the the background current is also shown as a black line.
 (c) Voltage-sustained self-oscillation observed in further devices.
}
\end{figure}

The solid black line in Fig.~4(a) is magnified and plotted over a larger voltage range in Fig.~4(b). It displays the measured DC current $\langle I_{\mathrm{SD}} \rangle$ of a shuttle that has not been triggered into self-oscillation. Clearly, no charge transport takes place even in the above-threshold regime of $\vert V_{\mathrm{DC}}\vert > 5$\,V. Figure~4(b) shows the background current over a voltage range extending to $\pm 60$\,V. It is fitted by a constant resistance of $R = 43$\,T$\Omega$ which is consistent with the leakage current through the wafer stack consisting of a $400$\,nm SiO$_2$ sacrificial layer and the silicon substrate.
The linear behavior reflects the unique ability of the nano-mechanical charge shuttles to withstand large electric fields without the onset of field emission or Fowler Nordheim tunneling~\cite{bib:Fowler1928} at larger bias voltages which often occurs between sharp-tipped nanoelectrodes~\cite{bib:Scheible2004b,bib:Ayari2007,bib:Kim2010b,bib:Weldon2010}.
We attribute the absence of field emission to the large vacuum distance of $2d = 140$\,nm between the source and drain electrode, along with their relatively large width (see Fig.~2(a) and (b)) preventing field enhancement. Thus we can conclude that the measured shuttling current can be purely attributed to the mechanical motion of the island over a voltage range exceeding $100$\,V.

Figure~4(c) shows further examples of voltage-sustained self-oscillation. The left panel was taken on the same array as Fig.~4(a), but acoustically triggered at $f=7.74$\,MHz with $35$\,dBm at $V_{\mathrm{SD}}=\pm 10$\,V, respectively.
The current-voltage characteristic in the right panel was measured on a different chip.
It displays a series of three sweeps performed in the order of the specified numbers, triggered at $f=8.98$\,MHz
with $26$\,dBm at $V_{\mathrm{SD}}=+10$\,V, $-14$\,V and $+14$\,V respectively,
indicating that the dissipation threshold can depend on the individual impact conditions. The observation that self-oscillation can be re-established several times with a new trigger is a generic feature of our shuttles. However, a trigger is essential to provide the shuttle with sufficient kinetic energy to effect island-electrode contact and thus to re-engage self-oscillation. The excess charge $\langle eN \rangle$ remaining on the island after a previous shuttling event is not sufficient to re-ignite self-oscillation upon increasing $V_{\mathrm{DC}}$ due to the large mechanical stiffness of the high stress SiN resonator.

The realization of voltage-induced self-oscillation in the high-temperature shuttling regime may open the pathway towards a second long sought goal in mechanical charge transport: Provided the charging energy of the island $E_{\mathrm{C}}=e^2/C_\Sigma$ with total island capacitance $C_\Sigma$ well exceeds the thermal energy $k_{\mathrm{B}}T$, charging of the island at the source or drain electrode becomes governed by Coulomb repulsion. In this so-called Coulomb blockade regime, only a well-defined number of electrons can enter the island such that the expectation value of the number of excess island charges $\langle N \rangle$ becomes an integer $N(V_{\mathrm{SD}})$, giving rise to a Coulomb staircase of discrete current steps~\cite{bib:Grabert1992}.

A major obstacle in reaching the low-temperature regime of discrete, Coulomb-blockade limited single electron shuttling with a piezo-driven device has so far been the accessible temperature range limited to above $10$\,K, whereas typical island capacitances $C_\Sigma$ of the order of $20$\,aF for $10 - 100$\,nm sized islands require lower temperatures
to observe clear Coulomb blockade. Voltage-sustained self-oscillation provides a minimum energy input scheme which should allow to lower the sample temperature by up to two orders of magnitude and operate the shuttle deeply in the Coulomb blockade regime. In order to facilitate millikelvin operation the inertial trigger required to induce self-sustained shuttling can be replaced by a capacitive trigger via an RF pulse applied between source and drain, as experimentally confirmed.

The observation of discrete single electron shuttling in the Coulomb blockade regime may even entail progress in metrology, where the realization of a quantum current standard would enable the testing of the metrological triangle~\cite{bib:Gallop2005}.
Using Ohm's law, a current given by the resistance and the voltage produced by the quantum Hall effect and the Josephson effect, respectively, can be tested against a metrological current source in order to check the consistency of the natural constants $e$ and $\hbar$. Unlike other proposed realizations~\cite{bib:Keller1996,bib:Pekola2008,bib:Steck2008}
a quantum current standard based on a single electron shuttle is not limited by coherent co-tunneling between source and drain.

Furthermore, the possibility of using ferromagnetic materials for the island as well as source and drain is a significant step towards the investigation of Kondo shuttling~\cite{bib:Kiselev2006}. The implementation of ferromagnetic materials may also lead to the realization of
spintronic devices such as mechanical spin valves~\cite{bib:Wang2008}.
Finally, a superconducting shuttle may allow to target mechanically mediating phase coherence in a nanostructured device~\cite{bib:Shekhter2006a}.\\

\begin{acknowledgments}
Financial support by the Future and Emerging Technologies programme of the European Commission, under the FET-Open project QNEMS (233992), the Deutsche Forschungsgemeinschaft via project Ko 416/18, the German Excellence Initiative via the Nanosystems Initiative Munich (NIM), LMUinnovativ as well as LMUexcellent is gratefully acknowledged. The authors appreciate ongoing support and stimulating discussion with J.~P.~Kotthaus. They also  thank C. Ecke for his assistance in taking the experimental data and S. Manus for expert technical help.
\end{acknowledgments}



%

\end{document}